\documentclass{article}[12pt]
\usepackage[dvips]{graphicx}

\begin{document}

\title{Numerical study of the spin-$1/2$ Heisenberg antiferromagnet 
on a 48-site triangular lattice \\
using the stochastic state selection method}

\author{Tomo Munehisa and Yasuko Munehisa \\ \\
Faculty of Engineering, University of Yamanashi, Kofu, Japan, 400-8511}




\maketitle

\begin{abstract}

We numerically study the magnetization and the dispersion relation 
of a frustrated quantum spin system. 
Our method, which is named the stochastic state selection method,
is a kind of Monte Carlo method to give eigenstates of the system 
through statistical averaging processes.

Using the stochastic state selection method with some constraints, 
we make a successful study of the spin-1/2 Heisenberg antiferromagnet on a  
48-site triangular lattice.
We calculate the sublattice magnetization and the static structure function in the 
ground state. 
Our result on the sublattice magnetization is 
consistent with the value given by the linear spin wave theory. 
This adds an evidence for the analysis based on the spontaneous symmetry 
breaking of the semi-classical N\'{e}el order in the ground state.

We also evaluate the low-lying one magnon spectra of the model 
with all wave vectors available on a 48-site triangular lattice. 
We find that at the ordering wave vector there is a Goldstone 
mode, which is in good agreement with the result from the spin wave analysis.
The magnon spectra with other wave vectors, however, are quite different from 
results obtained by the linear spin wave theory.
We observe a flat dispersion relation with a strong downward renormalization. 
Our results are compatible with those recently reported in the series expansion study 
and in the order $1/S$ calculation of the spin wave analysis.

\end{abstract}


\section{Introduction}
\label{sec1}

Since the pioneering work by Anderson and Fazekas\cite{Anderson, Anderson2}, 
the spin-1/2 Heisenberg antiferromagnet on a triangular lattice has been extensively
investigated as a promising candidate to realize a spin-liquid ground state
induced by geometric frustration and quantum fluctuations. 
Yet, in spite of a large amount of theoretical and experimental works,
we do not have any unified picture for this system.

On the theoretical side, most of the numerical studies carried out over the past 
decade with a variety of different techniques do not support that the suggested
spin-liquid ground state is realized in this model.
Instead they provide evidences 
to indicate the ground state with the three-sublattice order where the average direction of
neighboring spins differs by a $120^{\circ}$ angle\cite{Singh,Leung,
Sindz,Bernu,Boninsegni,Capri,Weihong,Trumper,ccm1,rev,Sorella2,ccm2,Sorella1,White3}.
Then the linear spin wave theory (LSWT)\cite{Trumper,Jolicoeur,Miyake,Chubukov,SpinWave} 
well describes numerical results calculated on lattices with finite sizes.

On the experimental side, several novel materials with
triangular structures have been investigated recently.
One of these materials is Cs$_2$CuCl$_4$ \cite{Exp5,Exp4}, which
is supposed to reduce to the one-dimensional spin-1/2 quantum  Heisenberg 
antiferromagnet because of its anisotropy\cite{QSL,Trumper2}. 
Other interesting materials  are
$\kappa$-(BEDT-TTF)$_2$Cu$_2$(CN)$_3$\cite{Exp1}
and EtMe$_3$Sb[Pb(dmit)$_2$]$_2$ \cite{Exp3,Exp2}, which are considered to be
close to the Heisenberg antiferromagnet on an isotropic triangular lattice.
These materials, however, do not show any magnetic long-range
order down to the quite low temperature compared with the exchange interactions.

Through further studies motivated by these experiments, theorists have found 
that fundamental properties on a triangular lattice are quite different 
from those on a square lattice, while antiferromagnets
on both lattices have the semi-classical long-range orders.
The dispersion relation is one of the properties that have been 
investigated to compare systems with different geometries. 
Recently the series expansion study\cite{Zheng2,Zheng} and 
the full 1/S calculation of the spin wave theory\cite{Chernyshev,Oleg,Chernyshev2} 
on this relation show that
on a triangular lattice one sees a downward renormalization of the higher
energy spectra, while on a square lattice one sees an upward renormalization.
The former authors also point out 
that the roton minimum is present in the relatively flat region of the 
dispersion relation on the triangular lattice.  
These features are quite different from the predictions of the LSWT.

In these somewhat confusing situations one needs unbiased numerical studies 
which do not depend on any special physical assumption.
The stochastic state selection (SSS) method, 
a new type of Monte Carlo method which we have developed  
in several years\cite{MM1,MMss,MM2,MM3,MMtri,MMeq,MMcnst}, 
has such a good property\footnote{All 
one assumes in numerical calculations is that the basis states have several general symmetries attributed to geometrical symmetries of the lattice on which the system lives.}.
One can therefore employ the method to evaluate any quantity in any system
one wants to investigate.

In the algorithm of the SSS method we consider the full Hilbert space of the system
and stochastically reduce it to relatively small one 
so that we can manage it in numerical studies. 
More concretely, we use a set of stochastic variables which are as many as 
basis states of the whole vector space under consideration, but
most of these variables are valued to be zero.
Then any inner product related to an arbitrary operator is calculable using the 
survived non-zero stochastic variables. Statistical averaging processes guarantee 
in a mathematically justified manner that the result becomes the correct value of 
the inner product.
It is found that several constraints on the set of stochastic variables 
are helpful to obtain better results with less statistical errors.   
Using this constrained SSS method we started our numerical study on the
spin-1/2 quantum Heisenberg antiferromagnet on a 48-site triangular lattice.
We have estimated lowest energy eigenvalues of the model
for each sectors with $0 \le S_z \le 4$, where $S_z$ denotes the $z$ component of the 
total spin $S$ of the system\cite{MMcnst}. 

In this paper we make a further investigation of the model 
by means of the constrained SSS method with two new applications.
One of them is to accurately calculate expectation values of 
operators which contain many off-diagonal elements in their representations. 
By evaluating the sublattice magnetization and the static structure function  
we demonstrate that it is possible 
to obtain accurate knowledge of the ground state in this method.
It should be noted that in the usual quantum Monte Carlo method 
these physical quantites are not easy to calculate even for non-frustrated systems.
Another is an extension to employ a new set of basis states 
with which complex coefficients are inevitable in an expansion of an arbitrary state. 
Using this set of basis states in the constrained SSS method we successfully 
calculate low-lying one magnon spectra with non-zero wave vectors.
It should also be noted that even for non-frustrated systems such as the
quantum Heisenberg antiferromagnet on a square lattice we cannot do without 
complex numbers in calculations with non-zero wave vectors.  
Our study in this paper performed by means of the constrained SSS method 
gives reliable results calculated from the first principle.
We see that our results are compatible with those in 
refs.\cite{Zheng2,Zheng,Chernyshev,Oleg,Chernyshev2}. 
It therefore supports the realization of an ordered ground state in the model.   
At the same time, however, it adds an evidence that 
dynamical properties of the system are not described by the LSWT.

The plan of this paper is as follows. In section \ref{sec2} we make brief 
descriptions of the model and the method.
Subsection \ref{subsec21} is to define the Hamiltonian of the model we study. 
In addition we comment on the power method. 
An operator related to the Hamiltonian is introduced here so that we can obtain 
the lowest eigenvalue of the Hamiltonian using the power method.
In subsection \ref{subsec22} the SSS method is shortly reviewed. Here 
an introduction of a two-valued stochastic variable that might be zero is essential.
Then the constrained SSS method is explained.     
In section \ref{sec3} we present our numerical results.
Subsection \ref{subsec31} is to show our definition and results on the sublattice magnetization and the static structure function. 
These results are compared with those reported on smaller lattices.
On the sublattice magnetization we also make a comparison with results calculated by 
the LSWT.
In subsection \ref{subsec32} we show our results on the one magnon spectra. 
For nine independent states in the Brillouin zone of a 48-site triangular lattice
we present estimated values of energy eigenvalues. 
Then the one-magnon dispersion relation is studied in comparison with results by
the LSWT. 
Section \ref{sec4} is devoted to summary.
Here we also count up characteristic features of our method.
Finally an appendix is given to explain our sets of basis states in detail.   

\section{Model and method}
\label{sec2}

\subsection{Model and power method}
\label{subsec21}

The Hamiltonian of the model on the $N$-site lattice is 
\begin{equation}
\hat H = J\sum_{(i,j)} \vec{\hat{S}}_i  \cdot \vec{\hat{S}}_j \,,
\label{hamil}
\end{equation}
where $\vec{\hat{S}}_i$ denotes the spin-1/2 operator on the $i$-th site  
and the sum runs over all $N_b(=3N)$ bonds of the $N$-site lattice. 
The coupling $J$ is set to 1 throughout this paper.
In our calculations we employ the power method. 
The SSS method is used to calculate expectation values of powers of an operator 
$\hat Q $,
\begin{equation} 
\hat Q \equiv l \hat I - \hat H ,
\label{qdef}
\end{equation}
where $\hat I$ denotes the identity operator and $l$ is a positive number. 
For each quantum number we choose one value of $l$ which ensures that the lowest  energy eigenvalue of the system corresponds 
to the eigenvalue of $\hat Q$ whose absolute value is the largest.
Detailed explanations are given in a previous paper\cite{MMcnst}.

In the power method we can estimate the energy eigenvalue $E$ by
\begin{equation}
 Q_{\rm E} \equiv l-E = \lim_{m\rightarrow \infty} \langle \psi_{\rm A} \mid \hat{Q}^m \mid 
\psi_{\rm A} \rangle / 
\langle \psi_{\rm A} \mid \hat{Q}^{m-1} \mid \psi_{\rm A} \rangle \,,
\label{qest}
\end{equation}
where $\mid \psi_{\rm A} \rangle$ denotes a state which approximates the exact eigenstate
$\mid \psi_{\rm E} \rangle$ of $\hat Q$,
\begin{equation}
 \mid \psi_{\rm A} \rangle = \mid \psi_{\rm E} \rangle c + \mid \zeta \rangle s 
\ \ \ (|s| \ll |c|) \, , 
\label{psiapx}
\end{equation}
since
\begin{equation}
\lim_{m\rightarrow \infty} \hat Q ^m \mid \psi_{\rm A} \rangle \propto \ 
\mid \psi_{\rm E} \rangle   
\label{psilim}
\end{equation}
holds.

We estimate an expectation value of an arbitrary operator $\hat O$ 
in the eigenstate $\mid \psi_{\rm E} \rangle$ using
\begin{equation}
 \langle \hat O \rangle 
= \lim_{m\rightarrow \infty} \langle \psi_{\rm A} \mid \hat O \hat{Q}^m \mid 
\psi_{\rm A} \rangle / \langle \psi_{\rm A} \mid \hat{Q}^m \mid \psi_{\rm A} \rangle \, .
\label{oest}
\end{equation}
Note that in this calculation a contamination appears at $s/c$ order because   
\begin{equation}
\langle \hat{O} \rangle =   
 \langle \psi_{\rm E} \mid \hat{O} \mid \psi_{\rm E} \rangle
  +  \langle \psi_{\rm E} \mid \hat{O} \mid \zeta \rangle s/c \, . 
\label{odiff}
\end{equation}

\subsection{Constrained SSS method}
\label{subsec22}

The stochastic state selection is realized by a number of stochastic variables.
Let us expand a normalized state ${\mid \psi \rangle}$ in an $N_{\rm V}$-dimensional 
vector space by a basis ${\{ \mid j \rangle \}}$,
\begin{equation} 
{\mid \psi \rangle} = {\sum_{j=1}^{N_{\rm V}} \mid j \rangle c_j}\, .
\label{expand}
\end{equation}
Then for each $j$ we generate a stochastic variable $\eta_j$ following to 
the on-off probability function,
%
\begin{equation}
P_j(\eta) \equiv 
\frac{1}{a_j}\delta(\eta -a_j) +(1- \frac{1}{a_j})\delta(\eta),  
\qquad \frac{1}{a_j} \equiv \min \left( 1, \frac{|c_j|}{\epsilon} \right) \, ,
\label{probf}
\end{equation}
with a positive parameter $\epsilon $ which is common to all ${P_j(\eta)}$ 
$(j=1,2,\cdots, N_{\rm V})$.  
Note that (\ref{probf}) implies $\eta_j=a_j \ (\geq 1)$ or $\eta_j=0$,
statistical averages being ${\langle \! \langle \eta_j \rangle \! \rangle} = 1$
and ${\langle \! \langle \eta_j^2 \rangle \! \rangle} = a_j$.
In numerical studies we can neglect basis states whose $\eta_j$=0. 
The parameter $\epsilon$ therefore controls the size of the partial vector space
we should consider in each sampling. Then calculations in the full vector space 
are to be recovered through statistical averaging processes. 

A random choice operator $\hat{M}$ is then defined by
%
\begin{equation}
\hat{M}\equiv \sum_{j=1}^{N_{\rm V}} \mid j \rangle \eta_j \langle j \mid \,.
\label{rchope}
\end{equation}
Using this $\hat{M}$ we obtain a state $\mid \tilde{\psi} \rangle$,
\begin{equation} 
\mid \tilde{\psi} \rangle \equiv \hat{M} {\mid \psi \rangle }= {\sum_{j=1}^{N_{\rm V}}
  \mid j \rangle c_j \eta_j},
\label{tildepsi} 
\end{equation}
which has less non-zero elements than ${\mid \psi \rangle}$ has.
It should be kept in mind that with any operator $\hat O$ the statistical average 
${\langle \! \langle \ \langle \psi_1 \mid \hat O \mid \tilde{\psi}_2 \rangle
\rangle \! \rangle }$ is exactly equal to the expectation value 
${\langle \psi_1 \mid \hat O \mid \psi_2 \rangle }$
because $\langle \! \langle \eta_j \rangle \! \rangle=1$ for all $j$. 

In calculations using the power method we generate independent random choice 
operators\footnote{Superscripts for random choice operators are added here to 
emphasize that for all $j$ the stochastic variable $\eta_j$ in $\hat{M}^{(k)}$ 
is generated independently to the one in $\hat{M}^{(l)}$ with $k \neq l$.}
$\hat{M}^{(1)} $, $\hat{M}^{(2)} $, $\cdots$ 
to obtain an improved state $\mid \psi_{\rm A}^{(n)} \rangle$  
from $\mid \psi_{\rm A} \rangle$.
Namely, we define $\mid \psi_{\rm A}^{(1)} \rangle \equiv \mid \psi_{\rm A} \rangle$ and 
\begin{equation}
 \mid \psi_{\rm A}^{(n)} 
\rangle \equiv \hat{Q} \hat{M}^{(n-1)}   \cdots \hat{Q} \hat{M}^{(1)}   \mid \psi_{\rm A} \rangle \ \ \ (n \geq 2) \, .
\label{psindef}
\end{equation}

In the constrained SSS method, we introduce $K$ ($K=1,2,\cdots$) constraints which can be
fulfilled by making $K$ $\eta_j$s depend on other $N-K$ independently generated stochastic
variables for each random choice operator. A detailed description on this improvement is given
in our previous paper\cite{MMcnst} together with numerical studies with a $K=1$ case. 
In the same manner as ref.\cite{MMcnst}, we use the state 
$\mid \phi_{\rm A}^{(n)} \rangle$  instead of the above mentioned 
$\mid \psi_{\rm A}^{(n)} \rangle$.
Here $\mid \phi_{\rm A}^{(n)} \rangle$ is defined by 
$\mid \phi_{\rm A}^{(1)} \rangle \equiv \mid \psi_{\rm A} \rangle$ and  
\begin{equation}
 \mid \phi_{\rm A}^{(n)} \rangle \equiv 
\hat{Q} \hat{M}_{\rm c}^{(n-1)}   \cdots \hat{Q} \hat{M}_{\rm c}^{(1)}   \mid \psi_{\rm A} \rangle \ \ \ (n \geq 2) \, ,
\label{phindef}
\end{equation} 
with random choice operators\footnote{Here we add a subscript in order to notify that 
each random choice operator is now determined to satisfy a given constraint.} 
$\hat{M}_{\rm c}^{(1)}$, $\hat{M}_{\rm c}^{(2)}$, $\cdots$, each of which is generated 
under a constraint 
\begin{equation}
\langle \psi_{\rm A} \mid \hat{M}_{\rm c}^{(k)} \mid \phi_{\rm A}^{(k)} \rangle
= \langle \psi_{\rm A} \mid \phi_{\rm A}^{(k)} \rangle \ \ \ (k=1,2, \cdots) \, .
\label{constcd}
\end{equation} 

We can evaluate an expectation value of an operator $\hat O$ 
by a statistical average of the ratio
\begin{equation}
R_{O}^{(m)} \equiv \frac{\langle \psi_{\rm A} \mid \hat O \mid \phi_{\rm A}^{(m)}
\rangle}{\langle \psi_{\rm A} \mid \phi_{\rm A}^{(m)} \rangle} \, ,
\label{ratioo}
\end{equation} 
with large enough value of $m$.
Actually, in order to save the computing time, either of the following two ratios is employed in our numerical study instead of (\ref{ratioo}) ,
\begin{equation}
R_{OM}^{(m)} \equiv \frac{\langle \psi_{\rm A} \mid \hat O \hat M_{\rm c}^{(m)}
\mid \phi_{\rm A}^{(m)}
\rangle}{\langle \psi_{\rm A} \mid \phi_{\rm A}^{(m)} \rangle} 
\label{ratioom}
\end{equation} 
or
\begin{equation}
R_{MO}^{(m)} \equiv \frac{\langle \psi_{\rm A} \mid \hat M'_{\rm c} 
\hat O \mid \phi_{\rm A}^{(m)}
\rangle}{\langle \psi_{\rm A} \mid \phi_{\rm A}^{(m)} \rangle} \, ,
\label{ratiomo}
\end{equation} 
where $\hat M'_{\rm c}$ denotes another random choice operator, which is generated 
independently to $\hat M_{\rm c}^{(1)}$ with the constraint 
$\langle \psi_{\rm A} \mid \hat{M}'_{\rm c} \mid \psi_{\rm A}^{(1)} \rangle
= \langle \psi_{\rm A} \mid \psi_{\rm A}^{(1)} \rangle$. 

It should be noted that the method is applicable for a state which has  
complex coefficients as well as a state with real coefficients. 
For a complex coefficient $c_j=|c_j|e^{i\theta_j}$ we obtain, 
after the stochastic selection, $c_j\eta_j=0$ or 
$c_j\eta_j=\epsilon e^{i\theta_j}$ if $|c_j| \leq \epsilon$.
With complex coefficients we newly have two technical difficulties. One of them 
is that we need double computer memory to treat complex numbers.
Another is that we cannot use the symmetries of rotation and reflection which reduce the number of independent basis states. 
   
\section{Results}
\label{sec3}

\subsection{Sublattice magnetization and static structure function }
\label{subsec31}

In this subsection we present our results of the sublattice magnetization and 
of the static structure function in the ground state, which we evaluate on 
the 48-site lattice using the ratio defined by (\ref{ratiomo}). 
We start from an approximate state constructed in the sector with $S_z=0$
since we want evaluate them in the ground state of the system. 
Detailed explanations of the way to calculate this approximate 
state are given in a previous paper\cite{MMcnst}.  

\begin{figure}[h]
\begin{center}
\scalebox{0.45}{\includegraphics{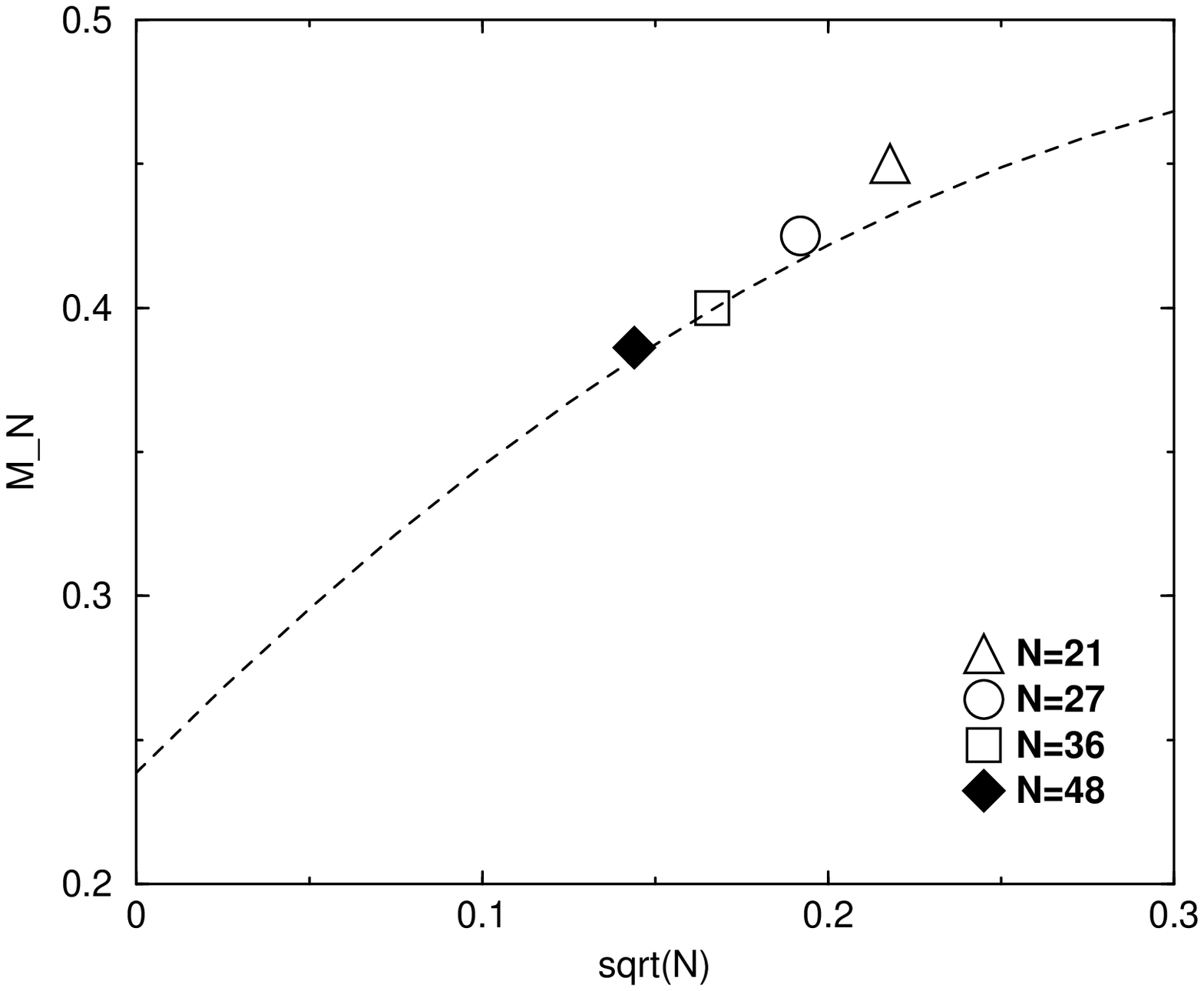}}
\caption{
Magnetization on the sublattice defined by $\sqrt{{M_N}^2}$. 
The filled diamond is our result on the 48-site lattice 
which we estimate by 50 samples of $R_{MO}^{(m)}$ in (\ref{ratiomo}) with   
$\epsilon=4 \times 10^{-3}$ and $m=15$, using the operator $\hat O$ in (\ref{osublmag2}).
The statistical error is much smaller than the mark. 
Data on smaller lattices\cite{Bernu2} and results from the LSWT\cite{Deutscher}, 
dashed line by (\ref{scallswt}), are also plotted.   
}
\label{fig:mag}
\end{center}
\end{figure}

Since the squared sublattice magnetization, which we denote by $M_N^2$ for the system 
with $N$ sites, is defined in ref.\cite{Bernu2} as 
\begin{equation}
M_N^2 = \langle \psi_{\rm E} \mid 
\frac{ {\frac{1}{3}} \left( \vec{\hat{S}}_{\rm A} ^2 + \vec{\hat{S}}_{\rm B} ^2 + 
\vec{\hat{S}}_{\rm C} ^2 \right) }
{4 \cdot \frac{N}{6} \left( \frac{N}{6} +1 \right)} \mid \psi_{\rm E} \rangle \, ,
\label{mnsqdef}
\end{equation}
the operator $\hat O$ in (\ref{ratioo}) for $M_N^2$ should be 
\begin{equation}
\hat O = \frac{ 3 \left( \vec{\hat{S}}_{\rm A} ^2 + \vec{\hat{S}}_{\rm B} ^2 + 
\vec{\hat{S}}_{\rm C} ^2 \right) }{ N \left( N +6 \right)} \, ,
\label{osublmag2}
\end{equation}
where A, B and C are sublattices of the $N$-site triangular lattice and
\begin{equation}
\vec{\hat{S}}_{\rm X} \equiv \sum_{i \in X} \vec{\hat{S}}_i \ \ \ 
\left(X= {\rm A}, \ {\rm B}, \ {\rm C} \right) \, .
\label{sxdef}
\end{equation} 
Our estimation of the sublattice magnetization with $N=48$ is $M_{48}= 0.380 \pm 0.05$. 
Fig.~\ref{fig:mag} plots this result together with those reported on $N=21$, 27 and 36 
lattices\cite{Bernu2}.
As is shown by calculations up to the 36-site lattice, 
the finite-size analysis on the magnetization suggests a finite value of the magnetization in the thermodynamic limit.
We see that our result is consistent with those for smaller lattices, 
which supports the scaling predicted by the LSWT,
\begin{equation}
M_N=a_0 + \frac{b_0}{\sqrt{N}} + \frac{c_0}{N} \, ,
\label{scallswt}
\end{equation}
where $a_0=0.2387$, $b_0=1.215$ and $c_0=-1.5$\cite{Deutscher}.
The finiteness of the sublattice magnetization, $M_{\rm \infty} > 0$, is also observed
in Fig.~\ref{fig:mag}. 

\begin{figure}[ht]
\begin{center}
\scalebox{0.45}{\includegraphics{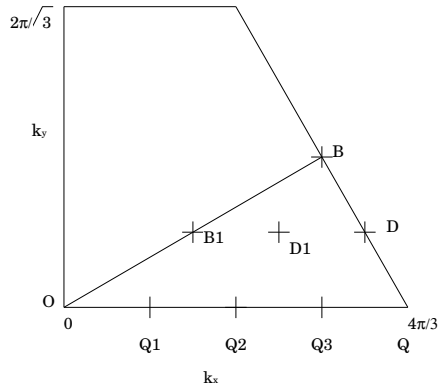}}
\caption{ Brillouin zone on the triangular lattice and the wave vectors
of the 48-site lattice.
}
\label{fig:Bz}
\end{center}
\end{figure}

\begin{figure}[h]
\begin{center}
\scalebox{0.45}{\includegraphics{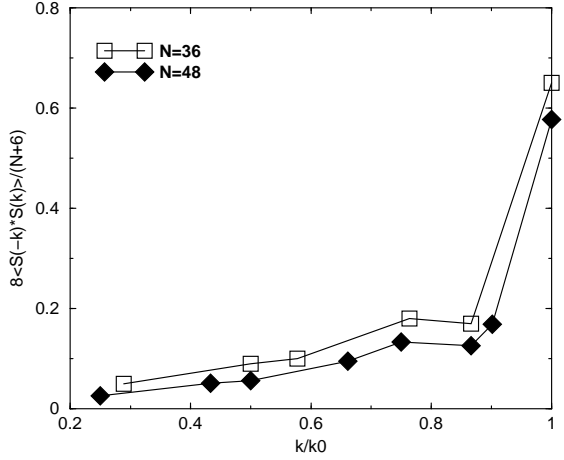}}
\caption{
Static structure function, $\langle {\vec{\hat S}}(-\vec{k}) \cdot \vec{\hat S}(\vec{k}) \rangle$ normalized by $8/(N+6)$ for $N$-site lattice. We obtain data on a 48-site lattice (filled diamonds) by $R_{MO}^{(m)}$ in (\ref{ratiomo}) with   
$\epsilon=5 \times 10^{-2}$ and $m=15$. Here the opertor $\hat O$ is given by (\ref{ostfunc})
and $k_0=4\pi/3$.
The number of samples is 50, with which statistical errors are quite small compared to marks.
Squares are data for the 36-site system from ref.\cite{Bernu}.
Solid lines are to guide eyes.
}
\label{fig:sf}
\end{center}
\end{figure}

Next we show results of the static structure function $S(\vec{k})$ in the ground state,
for which the operator $\hat O$ is defined by
\begin{equation}
\hat O = {\vec{\hat S}}(-\vec{k}) \cdot \vec{\hat S}(\vec{k}) 
\label{ostfunc}
\end{equation}
with 
\begin{equation}
\vec{\hat S}(\vec{k}) \equiv \frac{1}{\sqrt{N}}
\sum_i e^{i\vec{k} \cdot \vec{r}_i } \ \vec{\hat S}_i \, .
\label{skdef}
\end{equation}
Possible values of $|\vec{k}|/k_0$ on the 48-site lattice, where $k_0$=$4\pi/3$,
are obtained from the Brillouin zone shown in Fig.~\ref{fig:Bz}. They are  
$1/4$, $\sqrt{3}/4$, $1/2$, $\sqrt{7}/4$, $3/4$, $\sqrt{3}/2$, $\sqrt{13}/4$ and 1. 
Numerical results with a normalization factor $8/(N+6)$ 
are plotted in Fig.~\ref{fig:sf} as a function of $|\vec{k}|/k_0$.
In order to make a comparison between our data and previously reported ones for 
a smaller $N$, we also plot the 
structure function calculated on an $N=36$ lattice\cite{Bernu} in Fig.~\ref{fig:sf}.
We observe that in the full range of $|\vec{k}|/k_0$ 
our $N=48$ results lie slightly below those calculated on the 36-site lattice. 

\subsection{One magnon spectra}
\label{subsec32}

In this subsection we discuss the one magnon spectra with $S_z=1$.
Detailed descriptions of wave vectors $\vec{k}$ and a set of basis states 
$\{ \mid j (\vec{k}) \rangle \}$ we use here are given in the appendix. 

For each value of $\vec{k}=k_x \vec{e}_x + k_y \vec{e}_y$ 
on the 48-site lattice, we start from an approximate state 
$\mid \Psi_{\rm A} (\vec{k}) \rangle $ expanded by a set of basis states 
$\{ \mid j (\vec{k}) \rangle \}$ with complex coefficients,
\begin{equation}
\mid \Psi_{\rm A} (\vec{k}) \rangle = \sum_{j(\vec{k})=1}^{N_{\rm V}(\vec{k})} 
\mid j (\vec{k}) \rangle \, c_{j(\vec{k})} \, ,
\label{psikexpnd}
\end{equation}
which is used as $\mid \psi_{\rm A} \rangle$ in (\ref{phindef}), (\ref{constcd}) 
and so on.
The state $\mid \Psi_{\rm A} (\vec{k}) \rangle $ is prepared following descriptions in a previous paper\cite{MMcnst}.
In $\mid \Psi_{\rm A} (\vec{k}) \rangle $ we constructed for each $\vec{k}$, 
the number of basis states with non-zero $c_{j(\vec{k})}$, which we denote by $N_{\rm basis}$, ranges from 270 million to 1.6 billion.  
We evaluate an upper bound of the lowest energy eigenvalue $ \overline{E}(k_x,k_y)$, 
using the ratio $R_{OM}^{(m)} $ in (\ref{ratioom}) with the operator 
$\hat O = \hat Q = l\hat I - \hat H$. 
The results are presented in Table~\ref{tab:result} together with values of $l$
and $N_{\rm basis}$.

\begin{table*}[h]
\begin{center}
\begin{tabular}{|l|c|l|c|c|}   \hline
point & $(k_x,k_y)$ & $l$ & $N_{\rm basis}$&   $ \overline{E}(k_x,k_y)$ \\ \hline
O & $(0,0)$ &  5.2 &  $1.6 \times 10^9$ &   $-26.327 \pm 0.023$ \\ \hline
Q$_1$ &$(\pi/3,0) $ &  5.65 &  $4.0 \times 10^8$ &   $-25.413 \pm 0.048$ \\ \hline
Q$_2$ & $(2\pi/3,0) $&  5.8 & $3.7 \times 10^8$ &   $-25.202 \pm 0.065$ \\ \hline
Q$_3$ & $(\pi,0) $&  5.8 &  $3.5 \times 10^8$ &   $-25.241  \pm 0.061$ \\ \hline
Q & $(4\pi/3,0) $&  5.2 & $9.6 \times 10^8$ &   $-26.131 \pm 0.030$ \\ \hline
B$_1$ & $(\pi/2,\pi/(2\sqrt{3})) $&  5.8 & $3.2 \times 10^8$ & $-25.248 \pm 0.076$ \\ \hline
B & $(\pi,\pi/\sqrt{3}) $&  5.8 & $2.7 \times 10^8$ & $-25.419 \pm 0.062$ \\ \hline
D$_1$ & $(5\pi/6,\pi/(2\sqrt{3})) $&  5.6 & $3.6 \times 10^8$ & $-25.288 \pm 0.040$ \\ \hline
D & $(7\pi/6,\pi/(2\sqrt{3})) $&  5.65& $3.7 \times 10^8$ & $-25.416 \pm 0.050$ \\ \hline
\end{tabular}
\caption{Results for one magnon spectra for values of $\vec{k}$ at the points in the 
Brillouin zone shown in Fig.~\ref{fig:Bz}.
Here $N_{\rm basis}$ is the number 
of basis states we actually employ to construct the initial approximate state 
$\mid \Psi_{\rm A} (\vec{k}) \rangle $. 
The value of $l$ in $\hat Q$ defined by (\ref{qdef}) is determined 
so that it gives the lowest eigenvalue of $\hat H$\cite{MMcnst} with the given $\vec{k}$.
A typical value of $\epsilon$, the parameter in (\ref{probf}), is $0.002$, 
with which $\sim$ $1.3\%$ of the basis states survive after each stochastic selection.
Depending on $\vec{k}$, 
values of $m$ we employ to calculate the energy upper bound lie between 15 and 20, and
numbers of samples to calculate the average and the statistical error 
ranges between 100 and 200. } 
\label{tab:result}
\end{center}
\end{table*}   

\begin{figure}[h]
\begin{center}
\scalebox{0.45}{\includegraphics{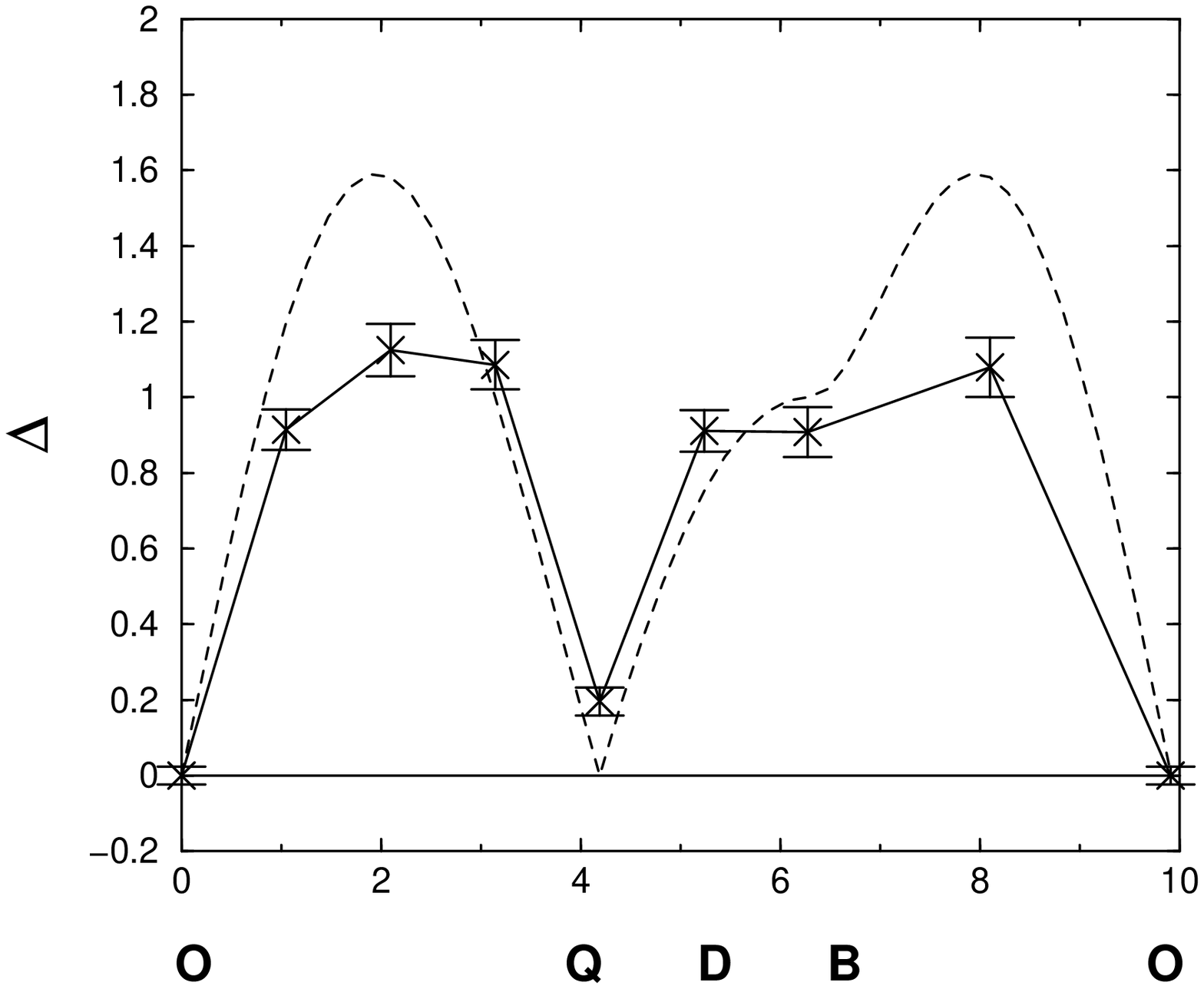}}
\caption{Magnon spectra along a path OQDBO in Fig.~\ref{fig:Bz}. 
Crosses are our data $\Delta (k_x,k_y)$ 
estimated on the 48-site lattice by one magnon gap energies (\ref{omaggpe}).
Statistical errors and a solid line to guide eyes are also presented in the figure.
The dashed line is the result from the LSWT\cite{SpinWave}. 
}
\label{fig:spectra}
\end{center}
\end{figure}

In Fig.~\ref{fig:spectra} we plot a gap energy of one magnon with a wave vector 
$(k_x,k_y)$, which is calculated from data in Table~\ref{tab:result} by
\begin{equation}
\Delta(k_x,k_y)= \overline{E}(k_x,k_y)-\overline{E}(0,0) \, ,
\label{omaggpe}
\end{equation}
along the path $OQDBO$ in Fig.~\ref{fig:Bz}. 
We also plot the dispersion relation calculated by the LSWT\cite{SpinWave} 
in Fig.~\ref{fig:spectra}.
We see that our data is consistent with the gapless state at the ordering wave vector, 
which is expected by the full breaking of the $SU(2)$ symmetry.
The gap energy at the central region of the Brillouin zone is much smaller than that 
calculated  by the LSWT analysis. 
This downward renormalization is in contrast with the upward renormalization 
found in the square lattice.
In the same central region we also find that the spectra is flat or dispersionless.

\vskip 0.5cm

\section{Summary}
\label{sec4}

In this paper we have reported further developments on numerical calculations  
by means of the stochastic state selection (SSS) method, 
which has the following good properties.
\begin{itemize}
\item With this method we can study large systems compared to the exact 
diagonalization\cite{Bernu}.
\item This method presents a solution for the so-called sign problem. Note that the
ordinary quantum Monte Carlo method, which has been quite powerful to study spin 
systems on bipartite lattices\cite{rev,book1,book2}, 
does not work in most frustrated systems.
\item One does not need any physical assumption. This is in contrast to the 
situation that recent variational approaches are based on some assumptions 
which are specific to the system to be studied.
For example, the coupled-cluster method\cite{ccm1,ccm2} is based on an eigenstate with 
a special form. In the density matrix renormalization group 
method \cite{White3,White1,White2,Hene,Mae}, on the other hand, 
one assumes that the ground state is represented by the matrix product state.
Another example is the stochastic reconfiguration method\cite{Sorella} 
whose origin is assigned to the fixed-node Monte Carlo method\cite{fixednode2,fixednode}.
In these methods an effective Hamiltonian is used instead of the exact one. 
\item In the SSS method it is easy to treat complex coefficients in the expansion of 
states. This is impossible in the ordinary Monte Carlo method.   
\end{itemize} 

Using this method we numerically study the spin-$1/2$ quantum  
Heisenberg antiferromagnet on the 48-site triangular lattice. We evaluate the sublattice magnetization and the static structure function in the ground state of the system. We also calculate the one magnon spectra with non-zero wave vectors.  

Our results on the magnetization and the structure function show that 
the SSS method is quite effective to estimate these quantities.
The accuracy of the data, 
which depends on how many samples are available to calculate
statistical averages, is satisfying within our computing resources. 
We observe that the results we obtain are consistent with the reported ones 
on smaller lattices. 
Therefore we see that the linear spin wave theory describes the results well 
so that an analysis based on the spontaneous symmetry breaking on the semi-classical
N\'{e}el order ground state is acceptable.

On the magnon spectra, we make sure that the one magnon energy at the ordering 
wave vector is consistent with the Goldstone mode, as is expected in the 
spin wave analysis. For other values of the wave vectors, however, 
we observe flat spectra with strong downward renormalization, which are quite
different from the spectra predicted by the LSWT.
We find our results are qualitatively consistent with results by the series
expansion study\cite{Zheng2,Zheng} and 
by the order $1/S$ calculation of the spin wave analysis\cite{Chernyshev,Oleg,Chernyshev2}.   
 
Finally let us again emphasize that 
we can get physical insight for the frustrated quantum spin systems 
from numerical results by the SSS method because the method is well developed 
to calculate various observables such as energy, the magnetization and the magnon spectra.
Especially it is notable that using this method we can evaluate the magnon spectra. 
While neither the ordinary Monte Carlo method nor the variational Monte Carlo method
is applicable to calculate this quantity, we can use the SSS method    
in the same manner as the one used to evaluate the ground state energy.

\vskip 1cm \noindent
{\Large {\bf Appendix}}

In this appendix we summarize our definition of wave vectors $\vec{k}$ and 
two sets of basis states $\{ \mid j \rangle \} $
and $\{ \mid j (\vec{k}) \rangle \}$ employed in this paper.
The former set $\{ \mid j \rangle \} $, which is introduced for the 
48-site lattice in the previous work with $\vec{k}= \vec{0}$~\cite{MMcnst}, 
is used to calculate the sublattice magnetization and the static structure function.
While the latter set $\{ \mid j (\vec{k}) \rangle \}$ is newly introduced here
in calculations of the one magnon spectra with non-zero wave vectors. 
 
Let us denote two basis vectors of the triangular lattice by $\vec{u}_1$ and $\vec{u}_2$, 
which are described by unit vectors $\vec{e}_x$ and $\vec{e}_y$ in $x$- and $y$-directions,
\begin{equation}
\vec{u}_1=\vec{e}_x \, , \ \ \ \vec{u}_2= \frac{1}{2} \vec{e}_x + \frac{\sqrt{3}}{2} \vec{e}_y \, .  
\label{u1u2}
\end{equation}
The wave vectors are defined by 
\begin{equation}
\vec{k} \equiv  \frac{K_1}{N} \vec{g}_1 + \frac{K_2}{N }\vec{g}_2 = k_x \vec{e}_x + k_y \vec{e}_y \, , \ \ \ (K_a=0,1,2,\cdots \ (a=1,2)) \, ,
\label{kdef}
\end{equation}
where $\vec{g}_a$ $(a=1$,2 $)$ are two basis vectors of the reciprocal lattice 
for which 
\begin{equation}
\vec{u}_a \cdot \vec{g}_b=2\pi \delta_{ab} \ \ \ (a=1,2, \ b=1,2) 
\label{uagb}
\end{equation}
holds.  

To construct a state with a fixed wave vector $\vec{k}$,
we first define translation operators $\hat{T}_a$ in the directions $\vec{u}_a$ $(a=1,2)$.
For any state $ \mid \Phi \rangle $ the translation operators 
$ \hat{T}_1$ and $ \hat{T}_2 $ produce new states if $ \mid \Phi \rangle $
is not an eigenstate of them. We define, with $n_a=1,2,\cdots$ $(a=1,2)$,  
\begin{equation}
\mid \Phi_{(n_1,n_2) } \rangle \equiv \hat{T}_1^{n_1}  \hat{T}_2^{n_2} \mid \Phi \rangle \, .
\label{phivrdef}
\end{equation} 
Using these states we make an eigenstate common to $ \hat{T}_1$ and $ \hat{T}_2 $,
\begin{equation}
\mid \Psi(\vec{k}) \rangle \equiv C
\sum_{n_1,n_2} \mid \Phi_{(n_1,n_2)} \rangle  
e^{-i \vec{k}\cdot (n_1\vec{u}_1 + n_2\vec{u}_2) }
\label{comeigens}
\end{equation}
Here the sum runs over all $n_1$ and $n_2$ for which 
$\mid \Phi_{(n_1,n_2)} \rangle$'s are linearly independent. 
The normalization factor $C$ in (\ref{comeigens}) is therefore $1/\sqrt{N}$ when all 
$ \mid \Phi_{(n_1,n_2)} \rangle   $ are independent to each other.
It is easy to see that 
\begin{equation}
\hat{T}_a  \mid \Psi (\vec{k}) \rangle =\mid \Psi(\vec{k}) \rangle 
e^{i \vec{k} \cdot \vec{u}_a}  \ \ \ (a=1,2) \, ,  
\label{treigenv}
\end{equation}
with the periodic boundary conditions in both directions.

In order to describe our sets of basis states we consider a state
$\mid s \rangle = \mid s^z_1, s^z_2, \cdots, s^z_N \rangle$  
having a fixed $z$ component $s^z_i= +1/2$ or $-1/2$ on each site $i$ of the lattice.
Then $2^N$ $\mid s \rangle$'s form a set of basis states $\{ \mid s \rangle \} $.  

The set of basis states $\{ \mid j \rangle \} $
is a complete set constructed from $\mid s \rangle  $'s 
for which the $z$-component of the total spin $S_z = \sum_i s^z_i =0$.  
Each state $\{ \mid j \rangle \} $ is defined by a linear combination of several 
$\mid s \rangle  $'s so that it has translational symmetries 
for the zero momentum and even under the $\pi$ rotation, $2\pi/3$ rotation and the 
reflection.  
Total number of the basis states, namely $N_{\rm V}$, in this set amounts $\sim$ $_N$C$_{N/2}/12N$ $ \sim $ $5.6 \times 10^{10}$ for the $N=48$ lattice\footnote{
Most of these basis states are linear combinations of $576(=12N)$ $\mid s \rangle$'s, but some 
contains $\mid s \rangle$'s less than $12N$.}.
     
The set of basis states 
$\{ \mid j (\vec{k}) \rangle \}$ should be constructed for each wave vector 
$\vec{k}$. 
A basis state $\mid j(\vec{k}) \rangle$ is determined by a state
$\mid s \rangle $ in the $S_z=1$ sector, using it as $ \mid \Phi \rangle $ 
in (\ref{phivrdef}).
Then the set of basis states $\{ \mid j(\vec{k}) \rangle \}$, which fulfil
\begin{equation}
\hat{T}_a  \mid j(\vec{k}) \rangle =\mid j(\vec{k}) \rangle 
e^{i \vec{k} \cdot \vec{u}_a}  \ \ \ (a=1,2) \, ,  
\label{treigenvj}
\end{equation}
is completed with as many states $\mid s \rangle$'s as to span 
the full Hilbert space with $S_z=1$. 
Note that with this set of basis states one should use complex coefficients in 
expansions of states in the calculations.
The total number of the basis states for the 48-site lattice is\footnote{
Here all basis states are constructed by $48(=N)$ $\mid s \rangle $'s.} 
$N_{\rm V}(\vec{k})= _{N}C_{N/2-1} / N$ $\sim $ $6.5 \times 10^{11}$.



\begin{thebibliography}{23}

\bibitem{Anderson} Anderson P W 1973 Matter. Res. Bull. {\bf 8} 153

\bibitem{Anderson2} Fazekas P and Anderson P W 1974 Philos. Mag. {\bf 30} 423

\bibitem{Singh}
Singh R R P and Huse D A 1992 Phys. Rev. Lett. {\bf 68} 1766

\bibitem{Leung}
Leung P W and Runge K J 1993 Phys. Rev. B {\bf 47} 5861

\bibitem{Sindz}
Sindzingre P, Lecheminant P and Lhuillier C 1994 Phys. Rev. B {\bf 50} 3108

\bibitem{Bernu}
Bernu B, Lecheminant P, Lhuillier C and Pierre L 1994 Phys. Rev.  B {\bf 50} 10048

\bibitem{Boninsegni}
Boninsegni M 1995 Phys. Rev. B {\bf 52} 15304

\bibitem{Capri}
Capriotti L, Trumper A E and Sorella S 1999 Phys. Rev. Lett. {\bf 82} 3899

\bibitem{Weihong}
Zheng W, McKenzie R H and Singh R R P 1999 Phys. Rev. B {\bf 59} 14367

\bibitem{Trumper}
Trumper A E, Capriotti L and Sorella S 2000 Phys. Rev. B {\bf 61} 11529

\bibitem{ccm1}
Farnell D J J, Bishop R F and Gernoth K A 2001 Phys. Rev. B {\bf 63} 220402

\bibitem{rev}
Richter J, Schulenburg J and Honecker A 2004 {\it Quantum Magnetism 
(Lecture notes in physics {\bf 645})}
ed U Schollw\"{o}ck, J Richter, D J J Farnell and R F Bishop 
(Berlin/Heidelberg: Springer-Verlag)

\bibitem{Sorella2}
Arrachea L, Capriotti L and Sorella S 2004 Phys. Rev. B {\bf 69} 224414

\bibitem{ccm2}
Kr\"{u}ger S E, Darradi R, Richter J and Farnell D J J 2006
Phys. Rev. B {\bf 73} 094404

\bibitem{Sorella1}
Yunoki S and Sorella S 2006 Phys. Rev. B {\bf 74} 014408

\bibitem{White3}
White S R and Chernyshev A L 2007 Phys. Rev. Lett. {\bf 99} 127004

\bibitem{Jolicoeur}
Jolicoeur Th and  Le Guillou J C 1989 Phys. Rev. B{\bf 40} 2727

\bibitem{Miyake}
Miyake S J 1992  J. Phys. Soc. Japan {\bf 61} 983

\bibitem{Chubukov}
Chubukov A V, Sachdev S and Senthil T 1994 J. Phys.: Condens. Matter {\bf 6} 8891

\bibitem{SpinWave}
Auerbach A 1998 
{\it  Interacting Electrons and Quantum Magnetism}
(Berlin: Springer) 

\bibitem{Exp5}
Coldea R, Tennant D A, Tsvelik A M and Tylczynski Z  2001 Phys. Rev. Lett. {\bf 86} 1335

\bibitem{Exp4}
Coldea R, Tennant D A and Tylczynski Z  2003 Phys. Rev. B {\bf 68} 134424 

\bibitem{QSL}
Balents L 2010 Nature {\bf 464} 199

\bibitem{Trumper2}
Trumper A E 1999 Phys. Rev. B {\bf 60} 2987

\bibitem{Exp1}
Shimizu Y, Miyagawa K, Kanoda K, Maesato M and Saito G 2003 Phys. Rev. Lett. {\bf 91} 107001

\bibitem{Exp3}
Itou T, Oyamada A, Maegawa S, Tamura M and Kato R 2007 J. Phys. :Condens. Matter  {\bf 19} 
145247

\bibitem{Exp2}
Itou T, Oyamada A, Maegawa S, Tamura M and Kato R 2008  Phys. Rev. B{\bf 77} 104413 

\bibitem{Zheng2}
Zheng W, Fjaerestad J O, Singh R R P, McKenzie R H and Coldea R 2006 Phys. Rev. Lett. {\bf 96} 057201

\bibitem{Zheng}
Zheng W, Fjaerestad J O, Singh R R P, McKenzie R H and Coldea R 2006 Phys. Rev. B {\bf 74} 224420

\bibitem{Chernyshev}
Chernyshev A L and Zhitomirsky M E 2006 Phys. Rev. Lett. {\bf 97} 207202

\bibitem{Oleg}
Starykh O A, Chubukov A V and Abanov A G 2006 Phys. Rev. B {\bf 74} 180403

\bibitem{Chernyshev2}
Chernyshev A L and Zhitomirsky M E 2009 Phys. Rev. B {\bf 79} 144416 

\bibitem{MM1}
Munehisa T and Munehisa Y 2003 J. Phys. Soc. Japan {\bf 72} 2759

\bibitem{MMss} 
Munehisa T and Munehisa Y 2004 J. Phys. Soc. Japan {\bf 73} 340

\bibitem{MM2} 
Munehisa T and Munehisa Y 2004 J. Phys. Soc. Japan {\bf 73} 2245

\bibitem{MM3} 
Munehisa T and Munehisa Y 2004 Numerical study for an equilibrium in the
recursive stochastic state selection method {\it Preprint} cond-mat/0403626 

\bibitem{MMtri}
Munehisa T and Munehisa Y 2006 J. Phys. : Condens. Matter {\bf 18} 2327

\bibitem{MMeq} 
Munehisa T and Munehisa Y 2007 J. Phys. : Condens. Matter {\bf 19} 196202

\bibitem{MMcnst} 
Munehisa T and Munehisa Y 2009 J. Phys. : Condens. Matter {\bf 21} 236008

\bibitem{Bernu2}
Bernu B, Lhuillier C and Pierre L 1992 Phys. Rev. Lett. {\bf 69} 2590

\bibitem{Deutscher} 
Deutscher R and Everts H U 1993 Z. Phys. B : Condensed Matter {\bf 93} 77

\bibitem{book1}
Hatano N and Suzuki M 1993
{\it Quantum Monte Carlo Methods in Condensed Matter Physics}
ed M Suzuki (Singapore: World Scientific) p~13 

\bibitem{book2}
De Raedt H and von der Linden W 1995  
{\it The Monte Carlo Method in Condensed Matter Physics}
ed K Binder (Berlin: Springer) p~249

\bibitem{White1}
White S R 1992 Phys. Rev. Lett. {\bf 69} 2863

\bibitem{White2}
White S R 1993 Phys. Rev. B {\bf 48} 10345

\bibitem{Hene}
Henelius P 1999 Phys. Rev. B {\bf 60} 9561

\bibitem{Mae}
Maeshima N, Hieida Y, Akutsu Y, Nishino T and Okunishi K 2001 Phys. Rev. E 
{\bf 64} 016705

\bibitem{Sorella}
Sorella S 2001 Phys. Rev. B {\bf 64} 024512

\bibitem{fixednode2}
van Bemmel H J M, ten Haaf D F B, van Saarloos W, van Leeuwen J M J
and An G 1994 Phys. Rev. Lett. {\bf 72} 2442

\bibitem{fixednode}
ten Haaf D F B, van Bemmel H J M, van Leeuwen J M J,
van Saarloos W and Ceperley D M 1995 Phys. Rev. B {\bf 51} 13039


\end{thebibliography}
\end{document}